\documentclass[prd,preprint,floatfix,amsmath,superscriptaddress,preprintnumbers,amssymb,showkeys,showpacs]{revtex4}
\usepackage{graphicx,dcolumn,bm}

\begin{document}
\input epsf.tex
\title{Energy  distribution in  Kerr-Newman space-time in Bergmann-Thomson formulation}
\author{ S.~S.~Xulu}
    \email[Email address : ]{ssxulu@pan.uzulu.ac.za}
    \affiliation{Department of  Computer Science,
                University of Zululand, Private Bag X1001,
                3886 Kwa-Dlangezwa,
                South Africa}
\date{\today}

\begin{abstract}
We obtain the energy distribution in the Kerr-Newman metric with
the help of Bergmann-Thomson energy-momentum complex. We find that
the energy-momentum definitions prescribed by   Einstein,
Landau-Lifshitz, Papapetrou, Weinberg, and Bergmann-Thomson give
the same and acceptable result and also support the  {\em
Cooperstock Hypothesis} for energy localization in general
relativity. The repulsive effect due to the electric charge and
rotation parameters of the metric is also reflected from the
energy distribution expression.
\end{abstract}

\pacs{04.20.Dw, 04.20.Cv, 04.20Jb. 04.70Bw }

\keywords{Bergmann-Thomson complex, Kerr-Newman black hole, and  Energy distribution}

\maketitle

\section{\label{sec:sec1}Introduction}

Energy-momentum is an important conserved quantity whose definition has been a focus of many
investigations. Unfortunately, there is still no generally agreed definition of energy and momentum
in general relativity (GR). This dilemma in GR is highlighted in an important paper by Penrose\cite{Penrose}
in the following way:
\begin{quotation}
``It is perhaps ironic that {\em energy conservation},
paradigmatic physical concept arising initially from Galileo's
(1638) studies of the motion of bodies under gravity, and which
now has found expression in the (covariant) equation \centerline{
$\nabla_a T^{ab} = 0  \hspace*{2.5in}(1)$} a cornerstone of
Einstein's (1915) general relativity---should nevertheless have
found no universally applicable formulation, within Einstein's
theory, incorporating the energy of gravity itself.''
\end{quotation}
Indeed, conservation laws of energy-momentum together with the
equivalence principle, played a significant role in guiding
Einstein's search for his generally covariant field equations. The
numerous attempts aimed at finding an expression for describing
energy-momentum distribution due to matter, nongravitational and
gravitational fields  gave rise to a large number of
energy-momentum complexes whose physical meaning have been
questioned (see \cite{LL,Papap,Weinberg,BT53,Moller58,Chandra}).

The absence of a generally accepted definition of energy
distribution in curved space-times has  led to doubts regarding
the idea of energy localization. Over the past two decades
considerable effort has been put in trying to define an
alternative concept of energy, the so-called quasi-local energy.
The idea in this case is to determine the effective energy of a
source by measurements on a two-surface. These masses are obtained
over a two-surface as opposed to an integral spanning over a
three-surface of a local density as is the case for
pseudocomplexes. To date a large number of definitions of
quasi-local mass have been proposed (see in Brown and York
\cite{BrownYork} and Hayward\cite{Hayward}).
Bergqvist\cite{Bergqv} considered quasi-local mass definitions of
Komar, Hawking, Penrose, Ludvigsen-Vickers, Bergqvist-Ludvigsen,
Kulkarni {\it  et al.}, and Dougan-Mason  and concluded
that no two of these definitions give agreed results for the Kerr
as well as  Reissner-Nordstr\"{o}m space-times. On the contrary, the  pioneering
work of Virbhadra and his collaborators (notably, Nathan Rosen of Einstein-Rosen gravitational waves fame), and others have
demonstrated with several examples that for a given spacetime, the
energy-momentum complexes of Einstein, Landau and Lifshitz,
Papapetrou, and Weinberg (ELLPW), show a high degree of
consistency in giving the same and acceptable energy and momentum
distributions (see for instance \cite{KSVpapers,KSVetal,Xulu3458,xulu6,Radinschi,Bringley,Gad,Vagenas}).
 Aguirregabiria \textit{et al}\cite{ACV96} showed that the ELLPW
energy-momentum complexes ``coincide''  for any metric of  Kerr-Schild
class.

The stationary axially symmetric and asymptotically flat
electrovac Kerr-Newman (KN) solution, characterized by  constant
parameters mass $M$, charge $Q$, and angular momentum $a$, is the
most general black hole solution to the Einstein-Maxwell
equations.   Virbhadra \cite{KSVprd} and  then following him
Cooperstock and Richardson \cite{CoopRi} showed (up to the third
order and seventh order, respectively, of rotation parameter) that
the energy-momentum complexes of Einstein and Landau-Lifshitz give
the same and reasonable energy distribution in KN space-time.
Further Aguirregabiria {\it et al.} \cite{ACV96} performed exact
computations for the energy distribution in KN space-time in
Kerr-Schild Cartesian coordinates. They showed that the energy
distribution in the prescriptions of Einstein, Landau-Lifshitz,
Papapetrou, and Weinberg (ELLPW) gave the same result. In this
paper we investigate the energy-momentum distribution of the
Kerr-Newman space-time using the Bergmann-Thomson prescription and
compare the results obtained by Aguirregabiria {\it et al.}


\section{\label{sec:sec2} Energy and momentum distributions in Bergmann-Thomson formulation  }

In this Section we first write the energy-momentum complex
formulated by Bergmann and Thomson and then  use it to compute
energy and momentum distributions in Kerr-Newman spacetime.

The energy-momentum complex of Bergmann-Thomson is\cite{BT53}

\begin{equation}
{\bf{B}}^{jk} = \frac{1}{16\pi}  {\cal B}^{jkl}_{\quad ,l}
\text{,} \label{Bjk}
\end{equation}
where
\begin{equation}
{\cal B}^{jkl} =  g^{ji} {\cal V}_i^{\  kl} \label{Kjkl}
\end{equation}
with
\begin{equation}
{\cal V}_i^{\  kl}\  =\ - {\cal V}_i^{\  lk}\ =\
\frac{g_{in}}{\sqrt{-g}}
         \left[-g \left( g^{kn} g^{lm} - g^{ln} g^{km}\right)\right]_{,m} \
. \label{Uikl}
\end{equation}
The Latin indices take values $0$ to $3$. The Bergmann-Thomson
energy-momentum complex satisfies the local  conservation laws
\begin{equation}
\frac{\partial {\bf{B}}^{jk}}{\partial x^k} = 0 \text{,}
\end{equation}
in any coordinate system; however, ${\bf{B}}^{jk}$ itself  does not
transform as a tensor under a general
coordinate transformation. The energy and energy current (momentum)
density components are respectively represented by
${\bf{B}}^{00}$ and ${\bf{B}}^{\alpha0}$.

Energy and momentum components $P^i$ are given by

\begin{equation}
  P^i = \int \int \int {\bf{B}}^{i0} dx^1 dx^2 dx^3 .
\end{equation}
Using Gauss's theorem one obtains energy $P^0$ and momentum
components $P^{\alpha}$ ($\alpha$ takes values $1$  to $3$) as
follows:

\begin{equation}
P^j = \frac{1}{16\pi} \int \int {\cal B}^{j0\alpha} n_{\alpha} ds
\text{.}
\end{equation}

The Kerr-Newman space-time in Kerr-Schild Cartesian coordinates
$\{t,x,y,z\}$ is expressed by the line-element\cite{DKS}:

\begin{eqnarray}
\small ds^2\  &=&\  dt^2 - dx^2 - dy^2 - dz^2 -
          \frac{2MR^3\ -\ q^2R^2}{R^4\ +\ a^2 z^2}\times\nonumber \\
&&\left[ dt\ +\ \frac{z}{R}\ dz\ +\ \frac{R}{R^2\ +\ a^2}
\left(x dx +y dy\right)+ \frac{a}{R^2+a^2} \left(x dy - y
dx\right) \right]^2 \text{,}\label{KNmetric}
\end{eqnarray}
where $R$ is defined by
\begin{equation}
\small \frac{x^2\ +\ y^2}{R^2\ + \ a^2}\ +\ \frac{z^2}{R^2}\
=\ 1 \ \ .
\end{equation}
Note that $R$ becomes the usual  spherical radial coordinate $r = \sqrt{x^2+y^2+z^2}$ for rotation
parameter $a=0$.

Now we use equations $(2), (3), (6), (7)$ and  $(8)$ to obtain
energy and momentum distributions in Kerr-Schild Cartesian
coordinates. In these coordinates, all the components of  the
metric tensor for the Kerr-Newman metric are nonvanishing and also
have lengthy expressions. It is extremely difficult to perform
these calculations manually. Therefore, we used {\em Mathematica
4.0}\cite{Mathematica} for computations. The  energy and momentum
components inside a 2-surface with constant $R$  are respectively
given by
\newpage
\begin{equation}
E(R)= M-\frac{q^2}{4 R} \left[ 1 + \frac{\left(a^2+R^2\right)} { a R } \arctan\left(\frac{a}{R}\right)\right]
\label{Enenrgy}
\end{equation}
and
\begin{equation}
 P_x(R) =  P_y(R) = P_z(R) = 0.
\label{Momentum}
\end{equation}

It is striking that Bergmann-Thomson complex gives the same result
as found by Aguirregabiria {\it et al.}\cite{ACV96}) who used
definitions  proposed by Einstein, Landau and Lifshitz,
Papapetrou, and Weinberg. The results found by us further support
the importance of the energy-momentum complexes, as these yield
consistent and meaningful results.

Now for our convenience we define
\begin{eqnarray}
{\cal E} :=  \frac{E}{M},   \nonumber \\
{\cal S}  :=   \frac{a}{M},   \nonumber \\
{\cal Q}  :=    \frac{q}{M},   \nonumber \\
{\cal R}  :=   \frac{R}{M}
\end{eqnarray}


\begin{figure*}
\epsfxsize 16cm
\epsfbox{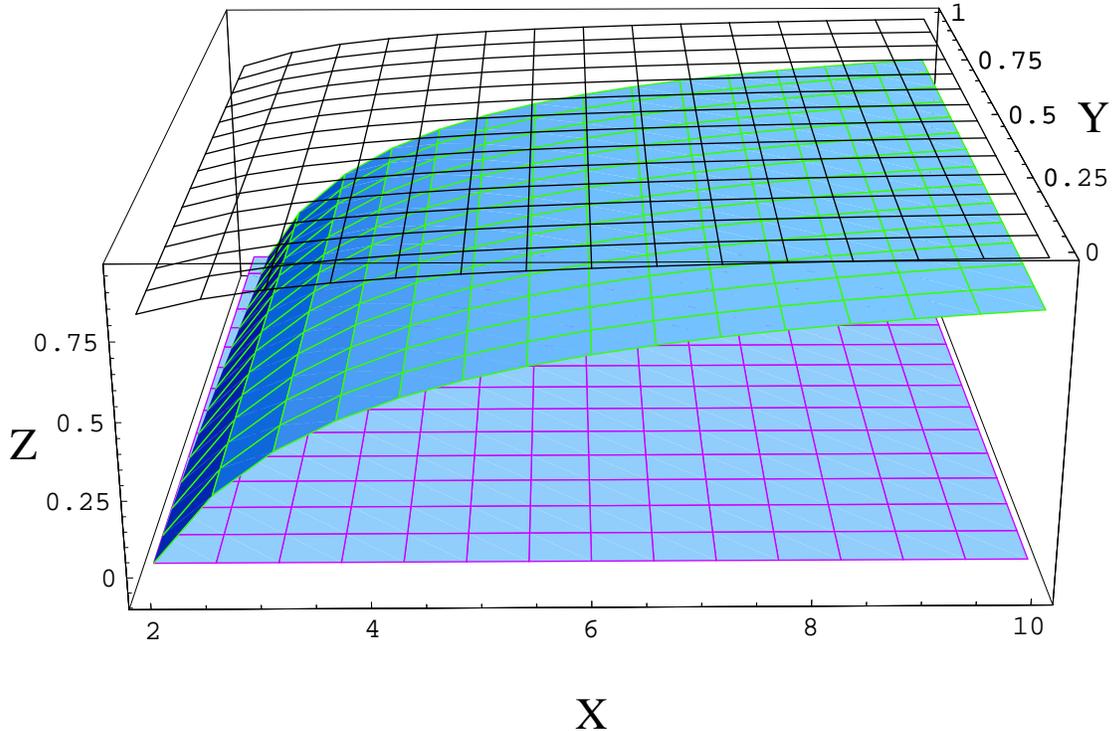}
\caption{
${\cal E}$ is plotted  on Z-axis  against  ${\cal R}$ on X-axis and ${\cal S}$ on Y-axis for
   ${\cal Q} = 0.9$ (top surface) and for  ${\cal Q} = 2.0$ (middle surface) . Momentum components which
   remain zero for all values of  ${\cal Q}, {\cal S}$ and  ${\cal R}$ are also shown by a flat surface.}
\label{fig1}
\end{figure*}
and plot ${\cal E}$ against ${\cal R}$  and  ${\cal S}$  for
different values of  ${\cal Q}$ (see Fig. 1). For the same value
of ${\cal R}$ and   ${\cal S}$,  ${\cal E}$ is higher for lower
values of  ${\cal Q}$. Linear momentum components $\{P_x, P_y,
P_z\}$ remain zero for all values of ${\cal Q},  {\cal S}$, and
${\cal R}$. This is also shown in the same figure.  Note that the
Kerr-Newman represents gravitational fields of black holes for
$M^2 \geq q^2 + a^2$ and of naked singularities for $M^2 < q^2 +
a^2$.


\section{\label{sec:sec3}Conclusion}

The main weakness of energy-momentum complexes is that
these restrict one to make calculations in `Cartesian
coordinates'. The alternative concept of quasi-local mass appears more
attractive because these are not restricted to the use of any special
coordinate system. There is a large number of quasi-local masses.
It has been shown\cite{Bergqv} that for the Kerr as well as Reissner-Nordstr\"{o}m space-times many
quasi-local mass formulations do not give agreed results. On the
other hand Aguirregabiria, Chamorro, and Virbhadra\cite{ACV96}
showed that the energy-momentum complexes of Einstein,
Landau-Lifshitz, Papapetrou, and Weinberg give the same result for
any metric of   Kerr-Schild class if the computations are carried
out in Kerr-Schild cartesian coordinates. The well-known
spacetimes of the Kerr-Schild class are for example the
Schwarzschild, Reissner-Nordstr\"{o}m, Kerr, Kerr-Newman, Vaidya,
Dybney \textit{et al.}, Kinnersley, Bonnor-Vaidya and
Vaidya-Patel. Virbhadra\cite{KSVprd} also showed that for a general
nonstatic spherically symmetric spacetime of the Kerr-Schild
class, the ELLPW give the same energy distribution as the Penrose
quasi-local mass definition if calculations are performed in
Kerr-Schild cartesian coordinates.

In the previous section we showed that the Bergmann-Thomson
energy-momentum complex gives the same energy and momentum
distribution in the Kerr-Newman spacetime as the ELLPW
prescriptions. These results and the important paper of Chang
\textit{et al}\cite{ChangEtAl}, which dispels doubts expressed
about the physical meaning of these energy-momentum complexes,
give confidence to the use of energy-momentum complexes to compute
energy-momentum distribution in a given spacetime.

The energy expression $E(R)$ in Eq. $(9)$ can be treated as the effective
gravitational mass that acts on a neutral (electric charge zero) test particle
situated at a coordinate distance $R$ from the centre of the Kerr-Newman black hole.
The repulsive effects of the electric charge and rotation parameters are obvious.
For sufficiently small $R$, $E(R)$ can be negative for large values of $q$ and $a$.
For Reissner-Nordstr\"{o}m metric, $E(r) < 0$ for $r < q^2/2M$.
As expected, the total energy ($\lim_{R \rightarrow \infty} E(r)$) of the Kerr-Newman metric comes to be the well-known ADM
mass $M$. It  is independent of the other two   (electric charge and rotation) parameters of the
charged rotating black hole.

A most  important implication of our result (expressed in Eq.
$(9)$) is that it supports {\em Cooperstock Hypothesis}
\cite{CoopHyp} for energy localization, which essentially states
that energy in a curved space-time is localized in the region of
nonvanishing energy-momentum tensor $T^{ik}$ of matter and
non-gravitational fields. For  the Kerr black hole space-time
$T^{ik}=0$ and therefore energy is confined to exterior of the
black hole; however,  for the Kerr-Newman black hole metric
$T^{ik} \neq 0$ and  the energy is distributed to the interior and
exterior of the black hole.
\acknowledgments I am very grateful to NRF of S. Africa for
financial support and Professor T. Dube  of the University of
Zululand for encouragement during the work.

\end{document}